\documentclass[conference]{IEEEtran}

\usepackage[pdftex]{graphicx}
\DeclareGraphicsExtensions{.pdf,.jpeg,.png}

\usepackage[cmex10]{amsmath}
\usepackage{mathabx}
\usepackage{algorithmic}
\usepackage{array}
\usepackage{mdwmath}
\usepackage{mdwtab}
\usepackage{eqparbox}
\usepackage{url}
\usepackage{hyperref}
\usepackage{csquotes}
\usepackage{array}

\usepackage{textgreek}
\usepackage{siunitx}
\usepackage{tikz}
\usepackage[utf8]{inputenc}
\usepackage{pgfplots}
\DeclareUnicodeCharacter{2212}{−}
\usepgfplotslibrary{groupplots,dateplot}
\usetikzlibrary{patterns,shapes.arrows}
\pgfplotsset{compat=newest}
\usepackage{booktabs}
\usepackage{multirow}
\usepackage{cite}

\begin{document}

\title{\LARGE Surrogate Parameters Optimization for Data and Model Fusion of COVID-19 Time-series Data}

\author{
\authorblockN{
Ogundare Timilehin
}
\authorblockA{Computer Science and Applied Mathematics, \\
 University of the Witwatersrand,\\
 Johannesburg, South Africa.
} \and
\authorblockN{
Terence L. van Zyl
}
\authorblockA{Institute for Intelligent Systems, \\
 University of Johannesburg,\\
 Johannesburg, South Africa.
}
}

\maketitle

\begin{abstract}
Our research focuses on developing a computational framework to simulate the transmission dynamics of COVID-19 pandemic. We examine the development of a system named ADRIANA for the simulation using South Africa as a case study. The design of ADRIANA system interconnects three sub-models to establish a computational technique to advise policy regarding lockdown measures to reduce the transmission pattern of COVID-19 in South Africa. Additionally, the output of the ADRIANA can be used by healthcare administration to predict peak demand time for resources needed to treat infected individuals. ABM is suited for our research experiment, but to prevent the computational constraints of using ABM-based framework for this research, we develop an SEIR compartmental model, a discrete event simulator, and an optimized surrogate model to form a system named ADRIANA. We also ensure that the surrogate's findings are accurate enough to provide optimal solutions. We use the Genetic Algorithm (GA) for the optimization by estimating the optimal hyperparameter configuration for the surrogate. We concluded this study by discussing the solutions presented by the ADRIANA system, which aligns with the primary goal of our study to present an optimal guide to lockdown policy by the government and resource management by the hospital administrators.
\end{abstract}

\IEEEoverridecommandlockouts

\IEEEpeerreviewmaketitle

% ===================
% # I. Introduction #
% ===================

\section{Introduction}
Towards the end of 2019, the world experienced the outbreak of coronavirus-2019 (COVID-19) pandemic which has had significant impact on everyday life~\cite{vanzyl2021did}. COVID-19 disease was identified as a novel disease that emanated from Wuhan, China \cite{cite1}. The World Health Organization (WHO) made a declaration on the 11\textsuperscript{th} of March 2020, that COVID-19 disease was a pandemic \cite{cite2}, as it was widespread globally.

Agent-Based Modeling (ABM) is an effective tool for simulating the discrete events of a pandemic \cite{cite38}. However, there are limitations to using an ABM approach for event simulation. These limitations include computational costs with resulting factors such as long simulation run times, high-level complexity of parameters, and complex model evaluation~\cite{cite39}.

Because of the constraints of ABMs mentioned above, we developed a surrogate model to substitute the ABM. We could minimize the burden of long simulation time and computational complexity by using the surrogate.

We use the Genetic Algorithm (GA) to parameterize the surrogate model~\cite{stander2020extended}. The GA offers problem-solving strategy by allowing us to determine the optimal hyperparameters for the surrogate. A study by Thengade \emph{et al.} \cite{cite3} confirms that the GA is a universal optimizer, which has been applied in various scientific disciplines and has been proven to perform well in the optimization of different machine learning models such as Radial Basis Function Networks (RBFNs) \cite{cite5}, Artificial Neural Networks (ANNs) \cite{cite7}, Kriging Model \cite{cite8,cite9} and Polynomial Regression \cite{cite10}. 

We develop a system named ADRIANA. The system alerts policymakers and hospital managements about the future transmission trend of COVID-19. The system will guide policymakers in determining when to increase or decrease the length of lock-downs, as well as recommend the duration of high resource demand for hospital management. ADRIANA is composed from three simulations:

\begin{itemize}
 \item The SEIR compartmental model;
 \item The COVID-19 discrete event simulator; and
 \item An optimized surrogate model.
\end{itemize}

The above-mentioned sub-models are discussed further below. 

In our study, we train and fit the surrogate with a historical COVID-19 dataset for South Africa. We implemented three deep neural network frameworks. They are Long Short-Term Memory (LSTM), Multi-Layer Perception (MLP), and Gated Recurrent Network (GRU), using LSTM as the benchmark model. We also implemented other machine learning architectures such as XGBRegressor (XGBR), Random Forest Regressor (RFR), Support Vector Regressor (SVR), Linear Regressing (LR), and Decision Tree Regressor (DTR). We evaluate each of the models with prediction-performance metrics, which are model Root Mean Squared Error (RMSE), Mean Absolute Error (MAE), and R-squared (R\textsuperscript{2}).

% =======================================================
% # II. Impact of traps on large signal characteristics #
% =======================================================

\section{Literature Review} \label{literature_review}
Since COVID-19 started, researchers across the globe have developed statistical and machine learning models to forecast the transmission rate of COVID-19. Despite advances in research, certain challenges persist, such as unstructured COVID-19 data and the use of statistical models to predict epidemic cases \cite{cite20,cite22}. Therefore, researchers are motivated to develop deep learning model to address these challenges and to predict the future trend of COVID-19 disease \cite{cite14,cite15,mathonsi2020prediction}.

To forecast a pandemic, the use of statistical models such as the Auto-Regressive (AR), Moving Average (MA) and, Integrated-ARMA (ARIMA) have been proven to be dependent on invalid assumptions~\cite{cite11,cite12}. In most cases, the statistical models are not able to fit the data properly, and evaluating their results proved difficult~\cite{cite11,cite12,cite13}. Previous studies forecasting COVID-19 were limited by assumptions made during model development~\cite{cite14,cite16}. The COVID-19 time-series data has non-stationary features, and using only epidemiological (such as reproduction number $R_{0}$) and statistical tools for modeling them can lead to incorrect predictions \cite{cite14,cite15,cite17}. Deep learning is less reliant on these assumptions, therefore it is worth exploring whether it is better at detecting patterns from COVID-19 dataset.

Additionally, previous prediction models did not consider patient recovery rate and the impact of government policy as factors in their models\cite{cite17}, this is captured in our simulation experiment. Our data exploration revealed that within the early stages of the pandemic, algorithms tend to allocate out-of-range hyperparameter values and neglect historical information.

Some of the existing models that forecasts the transmission trend of COVID-19 are CHIME \cite{cite18}, Adjutorium \cite{cite19}, Rationizer \cite{cite20}, and Epidemic Calculator \cite{cite21}. The researchers that developed these models were able to build a surrogate, train the surrogate, and make forecasts about the future transmission of COVID-19 in their respective regions. Some other synonymous study with specific focus are pandemic surge forecast by Andrew Crane-Droesch \emph{et al.} \cite{cite18}, scientific approach to pandemic control by Mihaela Van Der Schaar \emph{et al.} \cite{cite19}, simulation of pandemic transmission rate by Douglas \emph{et al.} \cite{cite20}, hospital capacity planning for epidemic by J. Ashleigh \emph{et al.} \emph{et al.} \cite{cite22}, and epidemic forecasting in Canada by Kumar \emph{et al.} \cite{cite23}.

% =============================================
% # III. Modeling and consistency validations #
% =============================================

\section{Methodology and Models} \label{methods_models}
The subsections that follow highlight the step-by-step approaches and workflow that we explored in our study.

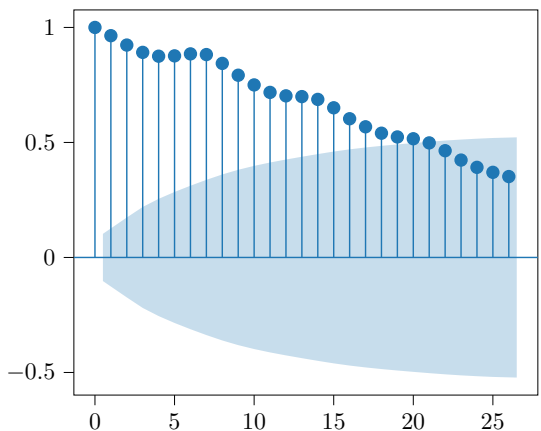
\begin{figure}[htb!] %!t
\centering
% This file was created by tikzplotlib v0.9.8.
\begin{tikzpicture}[scale=0.9]

\definecolor{color0}{rgb}{0.12156862745098,0.466666666666667,0.705882352941177}

\begin{axis}[
tick align=outside,
tick pos=left,
x grid style={white!69.0196078431373!black},
xmin=-1.325, xmax=27.825,
xtick style={color=black},
y grid style={white!69.0196078431373!black},
ymin=-0.598313096354929, ymax=1.07611014744547,
ytick style={color=black}
]
\path [fill=color0, fill opacity=0.25]
(axis cs:0.5,0.10273002627572)
--(axis cs:0.5,-0.102730026275721)
--(axis cs:2,-0.173708664947126)
--(axis cs:3,-0.219480810888443)
--(axis cs:4,-0.254854707981839)
--(axis cs:5,-0.284798508748047)
--(axis cs:6,-0.311967309286866)
--(axis cs:7,-0.337427207208406)
--(axis cs:8,-0.360941144751854)
--(axis cs:9,-0.38118389062554)
--(axis cs:10,-0.39818230560664)
--(axis cs:11,-0.412827029644975)
--(axis cs:12,-0.425788245504067)
--(axis cs:13,-0.437852162299808)
--(axis cs:14,-0.449484724509949)
--(axis cs:15,-0.460437336787745)
--(axis cs:16,-0.470041795405331)
--(axis cs:17,-0.47813768244657)
--(axis cs:18,-0.485207867655377)
--(axis cs:19,-0.491513180416466)
--(axis cs:20,-0.497371674748749)
--(axis cs:21,-0.502990964101673)
--(axis cs:22,-0.508169352651398)
--(axis cs:23,-0.5126226374279)
--(axis cs:24,-0.516299304347397)
--(axis cs:25,-0.51942738740356)
--(axis cs:26.5,-0.522202948909456)
--(axis cs:26.5,0.522202948909456)
--(axis cs:26.5,0.522202948909456)
--(axis cs:25,0.51942738740356)
--(axis cs:24,0.516299304347397)
--(axis cs:23,0.512622637427901)
--(axis cs:22,0.508169352651398)
--(axis cs:21,0.502990964101673)
--(axis cs:20,0.497371674748749)
--(axis cs:19,0.491513180416466)
--(axis cs:18,0.485207867655377)
--(axis cs:17,0.47813768244657)
--(axis cs:16,0.470041795405331)
--(axis cs:15,0.460437336787745)
--(axis cs:14,0.449484724509949)
--(axis cs:13,0.437852162299808)
--(axis cs:12,0.425788245504067)
--(axis cs:11,0.412827029644975)
--(axis cs:10,0.39818230560664)
--(axis cs:9,0.38118389062554)
--(axis cs:8,0.360941144751854)
--(axis cs:7,0.337427207208406)
--(axis cs:6,0.311967309286866)
--(axis cs:5,0.284798508748047)
--(axis cs:4,0.254854707981839)
--(axis cs:3,0.219480810888443)
--(axis cs:2,0.173708664947126)
--(axis cs:0.5,0.10273002627572)
--cycle;

\path [draw=color0, semithick]
(axis cs:0,0)
--(axis cs:0,1);

\path [draw=color0, semithick]
(axis cs:1,0)
--(axis cs:1,0.964163862586059);

\path [draw=color0, semithick]
(axis cs:2,0)
--(axis cs:2,0.923398537629198);

\path [draw=color0, semithick]
(axis cs:3,0)
--(axis cs:3,0.891603737278233);

\path [draw=color0, semithick]
(axis cs:4,0)
--(axis cs:4,0.874980652975843);

\path [draw=color0, semithick]
(axis cs:5,0)
--(axis cs:5,0.876445277628082);

\path [draw=color0, semithick]
(axis cs:6,0)
--(axis cs:6,0.885054960567496);

\path [draw=color0, semithick]
(axis cs:7,0)
--(axis cs:7,0.882048687796763);

\path [draw=color0, semithick]
(axis cs:8,0)
--(axis cs:8,0.843647173475596);

\path [draw=color0, semithick]
(axis cs:9,0)
--(axis cs:9,0.792250640929651);

\path [draw=color0, semithick]
(axis cs:10,0)
--(axis cs:10,0.750137930135013);

\path [draw=color0, semithick]
(axis cs:11,0)
--(axis cs:11,0.717615580108977);

\path [draw=color0, semithick]
(axis cs:12,0)
--(axis cs:12,0.702583985670388);

\path [draw=color0, semithick]
(axis cs:13,0)
--(axis cs:13,0.699309716704427);

\path [draw=color0, semithick]
(axis cs:14,0)
--(axis cs:14,0.687145263597002);

\path [draw=color0, semithick]
(axis cs:15,0)
--(axis cs:15,0.650694845750937);

\path [draw=color0, semithick]
(axis cs:16,0)
--(axis cs:16,0.603066353230036);

\path [draw=color0, semithick]
(axis cs:17,0)
--(axis cs:17,0.568059786305921);

\path [draw=color0, semithick]
(axis cs:18,0)
--(axis cs:18,0.540164692204211);

\path [draw=color0, semithick]
(axis cs:19,0)
--(axis cs:19,0.523906121438078);

\path [draw=color0, semithick]
(axis cs:20,0)
--(axis cs:20,0.516068121342327);

\path [draw=color0, semithick]
(axis cs:21,0)
--(axis cs:21,0.498075211980845);

\path [draw=color0, semithick]
(axis cs:22,0)
--(axis cs:22,0.464083859034952);

\path [draw=color0, semithick]
(axis cs:23,0)
--(axis cs:23,0.42335623582405);

\path [draw=color0, semithick]
(axis cs:24,0)
--(axis cs:24,0.391786371896078);

\path [draw=color0, semithick]
(axis cs:25,0)
--(axis cs:25,0.370100643894252);

\path [draw=color0, semithick]
(axis cs:26,0)
--(axis cs:26,0.351728661499211);

\addplot [semithick, color0]
table {%
-1.325 -1.11022302462516e-16
27.825 -1.11022302462516e-16
};
\addplot [semithick, color0, mark=*, mark size=2.5, mark options={solid}, only marks]
table {%
0 1
1 0.964163862586059
2 0.923398537629198
3 0.891603737278233
4 0.874980652975843
5 0.876445277628082
6 0.885054960567496
7 0.882048687796763
8 0.843647173475596
9 0.792250640929651
10 0.750137930135013
11 0.717615580108977
12 0.702583985670388
13 0.699309716704427
14 0.687145263597002
15 0.650694845750937
16 0.603066353230036
17 0.568059786305921
18 0.540164692204211
19 0.523906121438078
20 0.516068121342327
21 0.498075211980845
22 0.464083859034952
23 0.42335623582405
24 0.391786371896078
25 0.370100643894252
26 0.351728661499211
};
\end{axis}

\end{tikzpicture}
\caption{ACF-Plot Representation of the CRD Dataset}
\label{fig:acf_confirmed}
\end{figure}

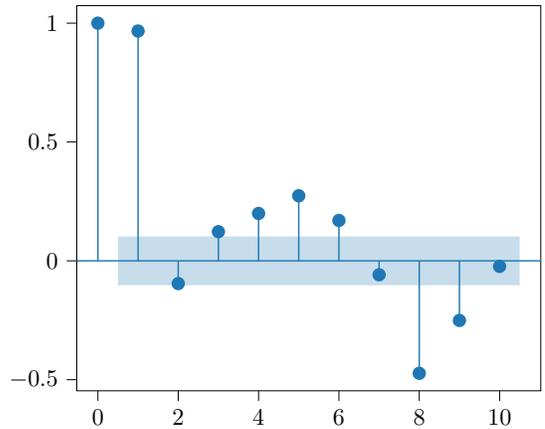
\begin{figure}[htb!]
  \centering
  % This file was created by tikzplotlib v0.9.8.
\begin{tikzpicture}[scale=0.9]

\definecolor{color0}{rgb}{0.12156862745098,0.466666666666667,0.705882352941177}

\begin{axis}[
tick align=outside,
tick pos=left,
x grid style={white!69.0196078431373!black},
xmin=-0.525, xmax=11.025,
xtick style={color=black},
y grid style={white!69.0196078431373!black},
ymin=-0.546740899123942, ymax=1.07365432852971,
ytick style={color=black}
]
\path [fill=color0, fill opacity=0.25]
(axis cs:0.5,0.102730026275721)
--(axis cs:0.5,-0.102730026275721)
--(axis cs:2,-0.102730026275721)
--(axis cs:3,-0.102730026275721)
--(axis cs:4,-0.102730026275721)
--(axis cs:5,-0.102730026275721)
--(axis cs:6,-0.102730026275721)
--(axis cs:7,-0.102730026275721)
--(axis cs:8,-0.102730026275721)
--(axis cs:9,-0.102730026275721)
--(axis cs:10.5,-0.102730026275721)
--(axis cs:10.5,0.102730026275721)
--(axis cs:10.5,0.102730026275721)
--(axis cs:9,0.102730026275721)
--(axis cs:8,0.102730026275721)
--(axis cs:7,0.102730026275721)
--(axis cs:6,0.102730026275721)
--(axis cs:5,0.102730026275721)
--(axis cs:4,0.102730026275721)
--(axis cs:3,0.102730026275721)
--(axis cs:2,0.102730026275721)
--(axis cs:0.5,0.102730026275721)
--cycle;

\path [draw=color0, semithick]
(axis cs:0,0)
--(axis cs:0,1);

\path [draw=color0, semithick]
(axis cs:1,0)
--(axis cs:1,0.966819961381062);

\path [draw=color0, semithick]
(axis cs:2,0)
--(axis cs:2,-0.0956287285792114);

\path [draw=color0, semithick]
(axis cs:3,0)
--(axis cs:3,0.122656201025305);

\path [draw=color0, semithick]
(axis cs:4,0)
--(axis cs:4,0.199353027713091);

\path [draw=color0, semithick]
(axis cs:5,0)
--(axis cs:5,0.273619733175885);

\path [draw=color0, semithick]
(axis cs:6,0)
--(axis cs:6,0.169895950623831);

\path [draw=color0, semithick]
(axis cs:7,0)
--(axis cs:7,-0.0587058288061787);

\path [draw=color0, semithick]
(axis cs:8,0)
--(axis cs:8,-0.473086570594231);

\path [draw=color0, semithick]
(axis cs:9,0)
--(axis cs:9,-0.250730939830534);

\path [draw=color0, semithick]
(axis cs:10,0)
--(axis cs:10,-0.0233166732407407);

\addplot [semithick, color0]
table {%
-0.525 -1.11022302462516e-16
11.025 -1.11022302462516e-16
};
\addplot [semithick, color0, mark=*, mark size=2.5, mark options={solid}, only marks]
table {%
0 1
1 0.966819961381062
2 -0.0956287285792114
3 0.122656201025305
4 0.199353027713091
5 0.273619733175885
6 0.169895950623831
7 -0.0587058288061787
8 -0.473086570594231
9 -0.250730939830534
10 -0.0233166732407407
};
\end{axis}

\end{tikzpicture}
  \caption{10-Lags PACF Plot Representation of CRD Dataset}
  \label{fig:pacf_confirmed}
\end{figure}

\subsection{Data Collection and Pre-processing}
The datasets used for our study were extracted from the COVID-19 Data Repository by the Center for Systems Science and Engineering (CSSE) at Johns Hopkins University \cite{cite24}. The dataset is presented in a time-series format which contains the daily cumulative statistics of COVID-19 confirmed cases, recovered cases, and death cases (CRD) for 191 countries across the globe starting from January $22^{nd}$, 2020 up till day$(X{t}-1)$, where X\textsubscript{t} is the current day. We then derived the daily instance for day($n$) by deducting the cumulative value of day$(n-1)$ from the cumulative value of day$(n)$.

To build our surrogate model with the datasets, we test for stationarity. The stationarity of the time-series datasets transforms it to a constant mean, constant variance, and constant probability distribution at any chosen point $t_{0} - t_{n}$ \cite{cite25} in the series. In addition, because unwanted patterns in the dataset are removed, the stationarity of the CRD dataset offers a higher level of certainty in the simulation result.

To verify for the stationarity of the CRD dataset, we plotted each of the CRD datasets on an ACF plot. Figure~\ref{fig:acf_confirmed} shows the ACF-plot for the confirmed cases dataset and we observed a similar trend of non-stationarity in the recovered and death datasets \cite{cite25}.

In addition, we used the Augmented Dickey-Fuller (ADF) statistical test to evaluate if the unit root exists in the CRD dataset. By default, ADF-test presents the null hypothesis that our dataset has a unit root \cite{cite28}, showing that our CRD dataset is not stationary.

The p-value estimated by the ADF-test for confirmed, recovered and death cases are 0.12, 0.98, and 0.96 respectively, which indicates that we can accept the null hypothesis that the CRD datasets are not stationary. Therefore, we apply the differencing technique to transform the CRD datasets into a stationary series.

By analyzing the PACF-plot in Figure~\ref{fig:pacf_confirmed}, we observe that differencing of order $1$ (i.e. $d=1$ ) is sufficient to transform each of the CRD datasets to a stationary series. 

Additionally, we normalize the CRD datasets before training and fitting the surrogate by re-scaling the data to a range between 0 and 1. We accomplished this to enhance the surrogate's learning rate and prediction accuracy during model training. 

\subsection{SEIR Compartmental Model}

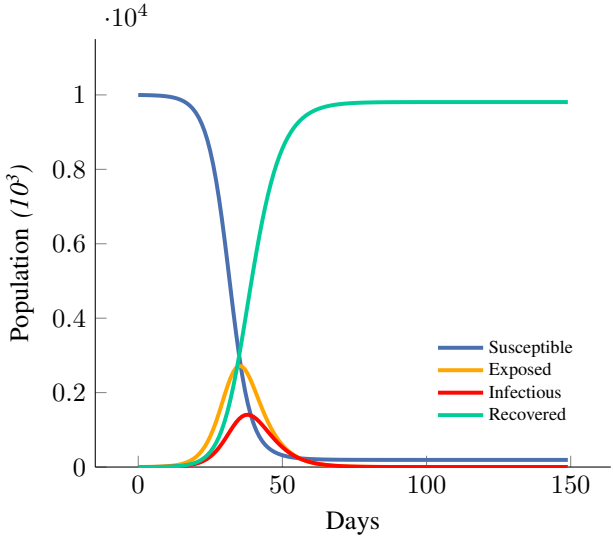
\begin{figure}[htb!]
  \centering
  \begin{tikzpicture}[scale=1.0]

% \definecolor{color2}{rgb}{0.866666666666667,0.517647058823529,0.32156862745098}
\definecolor{dblue}{rgb}{0.29,0.44,0.69}
\definecolor{dred}{rgb}{1.0,0.03,0.0}
\definecolor{dyellow}{rgb}{1.0, 0.65 0.0}
\definecolor{dgreen}{rgb}{0.0, 0.8,0.6}

\begin{axis}
[
legend cell align={left},
axis x line*=bottom,
axis y line*=right,
xtick pos=bottom,
ytick pos=left,
legend style={
    at={(0.95,0.2)},anchor= east,
    nodes={scale=0.69},
    draw=none
},
xlabel={Days}, ylabel={Population \emph{(10\textsuperscript{3})} }, 
ymin= 0, ymax= 11500
]

\addplot [line width=1.5pt,,dblue]
table {./tikz/b_susceptible.txt};
\addlegendentry{Susceptible}

\addplot [line width=1.5pt,,dyellow]
table {./tikz/b_exposed.txt};
\addlegendentry{Exposed}

\addplot [line width=1.5pt,,dred]
table {./tikz/b_infectious.txt};
\addlegendentry{Infectious}

\addplot [line width=1.5pt,,dgreen]
table {./tikz/b_recovered.txt};
\addlegendentry{Recovered}

\end{axis}
\end{tikzpicture}
  \caption{Numerical Simulation of SEIR Compartmental Model for COVID-19 in South Africa}
  \label{fig:seir_simulation}
\end{figure}

The Susceptible, Exposed, Infectious, and Recovered (SEIR) compartmental model is a mathematical model that was developed by mathematics-of-epidemiology experts McKendrick and Kermack for grouping homogenous populations into different SEIR compartments during pandemic outbreak \cite{cite29}. The SEIR model is used to provide computational insight into the dynamics that are involved in the spread of pandemic disease based on observed local factors such as disease prevalence rate and disease transmission pattern. 

In this study, we develop the SEIR model using Euler's method of mathematical differential equation \cite{cite30}. We simulate the number of people that recovered or died from the disease based on factors such as individual risk score, age, disease historical record, etc. Equation~\ref{eqn:seir_equation} shows the mathematical simulation of grouping population into the susceptible, exposed, infectious and recovered compartments respectively over a defined period $t$ during a pandemic. 

During model development, we considered internal factors contributing to the SEIR model architecture. Factors include initial disease prevalence rate, population size, disease transmission rate, incubation period, recovery rate, duration of infection, and reproduction number. Generally, the SEIR model offers insight into categorizing South Africa populations into epidemic compartments, as illustrated in Figure~\ref{fig:seir_simulation}. The proposed SEIR model takes the form:
\begin{equation}
\begin{cases}
S^{'}(t)  = -\frac{\beta SI}{N}  & \quad \text{Susceptible compartment } \\
E^{'}(t)  = \frac{\beta SI}{N} - \sigma E  & \quad \text{Exposed compartment } \\
I^{'}(t)  = \sigma E - \gamma I  & \quad \text{Infectious compartment } \\
R^{'}(t)  = \gamma I  & \quad \text{Recovered compartment} \\
\end{cases}
\label{eqn:seir_equation}
\end{equation}
where $N = S(t) + E(t) + I(t) + R(t)$ is the total population to be grouped into compartments.
The internal parameters of the SEIR model includes infection rate (\textbeta), incubation rate (\textsigma), incubation period (1/\textsigma), recovery rate (\textgamma), and reproduction number (R\textsubscript{o}) = $\frac{\beta}{\gamma}$.

\subsection{Discrete Event Simulation}
As part of our research method and architecture for the ADRIANA system, we developed a Discrete Event Simulation (DES) using SimPy \footnote{Available at https://simpy.readthedocs.io/en/latest/} to demonstrate how individuals in the exposed compartment of the SEIR model move to the infectious stage and their response to treatment which determines if an infected individual recovered or died during hospital treatment. We simulate patient activity based on defined model variables and individual personality traits, such as time of patient arrival, disease severity level, duration of treatment, time of exit from the hospital, age group, etc. This is comparable with range of population parameters that were enumerated by the World Health Organization (WHO)\footnote{https://www.afro.who.int/} and the National Institute of Communicable Diseases (NICD)\footnote{https://www.nicd.ac.za/} in South Africa. 

The architecture of the discrete event simulation presupposes that the simulated system changes at distinct points in simulation time. The simulation system is asynchronous, which implies that it does not operate on a perpetual clock but on random simulation time intervals \cite{cite40}. Furthermore, we use the DES to simulate the social processes that infected individuals undergo during a pandemic, with a starting population sample $N$ estimated as the number of infectious population in the SEIR model.

\subsection{Surrogate Model and Evaluation}
Recurrent Neural Network (RNN) performed well in modeling sequential data \cite{cite31}. Generally, RNN gives an accurate prediction of time-series sequential data which may be difficult to model with other computational algorithms. Our study implemented three neural network models which are LSTM, GRU, and multi-layer perceptron (MLP). 

\subsubsection{Long Short-Term Memory (LSTM)}
The LSTM is an archetype of the RNN that can learn long-term dependencies and remember past information to predict into an extended period \emph{t} in the future. LSTM is widely used in modeling because of its reliable results in solving varieties of scientific problems. LSTM has three major gates in its architecture. They are input gate, forget gate, and output gate. In our study, we implemented one input layer, three hidden layers, and one output layer for LSTM. We used Distributed Evolutionary Algorithm in Python (DEAP)\footnote{Available at https://deap.readthedocs.io/ and http://github.com/DEAP/deap} as the optimization engine, by minimizing the Root Mean Squared Error (RMSE) as the loss function in our optimization. Our optimization proposed the best number of LSTM neurons to be 58, and the best window size to be 9. We used \emph{ReLU} as activation function and \emph{ADAM} as internal model optimizer for LSTM. To avoid overfitting the LSTM model, we introduced a dropout layer and early stopping criteria. The LSTM surrogate implementation in our study has 50,954 trainable internal parameters. More knowledge on the LSTM framework can be accessed at \cite{cite32}.

\subsubsection{Gated Recurrent Units (GRU)}
The Gated Recurrent Units (GRU) has a similar network architecture with LSTM. Contrary to LSTM, GRU has only two gates which are the reset gate and update gate. GRU has been proven to be less complex in its operation and quicker for training and execution \cite{cite31,cite32}. This inspired us to use GRU into our study. The GRU surrogate implementation in our study has 38,734 trainable internal parameters.

\subsubsection{Multi-layer Perceptron (MLP)}
In this study, we used the back-propagation technique to train a multi-layer feedforward neural network (FFNN). MLP FFNN has an input layer, at least one hidden layer and an output layer \cite{cite31}.

The COVID-19 CRD datasets are a time-series format of data points on a 24-hours interval. We also model our datasets with the statistical autoregression AR and ARIMA model. This was done to compare the performance of the neural network models to the statistical regression models using the appropriate evaluation metrics. Table~\ref{tab:model_evaluation_metrics} shows the performance rank of the models considered in this study. We trained our surrogates and search for optimized hyperparameters by using 10-fold cross-validation.

\subsection{Model Performance Metrics}
The performance evaluation of a surrogate model is determined by how accurately it can relate the input parameters to the output sequence \cite{cite34}. We explore three performance metrics to evaluate the prediction of the surrogate model. They are Mean Absolute Error (MAE), R-squared (R\textsuperscript{2}), and Root Mean Squared Error (RMSE). MAE is calculated by summing the absolute difference between the actual value and the predicted value of the model. 

$R^{2}$ estimates the degree of similarity of data points to a fitted regression line. It indicates the interdependence level of variables in the same computation environment. \cite{cite35}. The $R^{2}$ accuracy is estimated by observing the range of its value between 0 and 1. Generally, $R^{2}$ value close to 1 suggests a better fit of the model, indicating a better correlation between the actual and predicted values.

Root Mean Square Error (RMSE) measures the standard deviation of model prediction errors by showing the distance between a regression line and data points \cite{cite36}. It generally describes the level of data concentration around an optimal fit. RMSE is widely used as an evaluation metric in forecasting, climatology, and regression analysis \cite{cite36}. 

\subsection{Surrogate Optimization}
It is essential to optimize the surrogate models. The optimization will allow us to search, identify, and configure the optimal surrogate hyperparameters. We employed the Genetic Algorithm (GA) as our surrogate optimizer because of its wide spectrum of uses in optimization problems across different scientific modeling fields \cite{cite4}.

\subsubsection{Hyperparameter Selection}

\begin{table}[htb!]
\caption{Hyperparameter Selection}
\label{tab:hyperparameter_selection}
\centering
\renewcommand{\arraystretch}{1}
\begin{tabular}{ l|lrrr }
\toprule
\textbf{Model} & \textbf{Hyperparameter} & \textbf{C} & \textbf{R} & \textbf{D} \\ 
\bottomrule
\toprule
Deep  & Neurons & 58 & 58 & 58 \\
Learning  & Window Size & 9 & 9 & 9 \\ \midrule
XGBR & Learning Rate & 1 & 0.1 & 0.1 \\ 
  & Max Depth & 1 & 1 & 1 \\ 
  & Max features & 10 & 10 & 10 \\
  & Estimators & 25 & 25 & 25 \\ \midrule
SVR & Epsilon & 0.6 & 0.6 & 0.4 \\ 
    & Kernel & Linear & Linear & Linear \\ \midrule  
DTR  & Max Depth & 4 & 4 & 4 \\ 
    & Min Split & 7 & 7 & 7 \\
    & Min Leaf & 3 & 3 & 3 \\   \midrule 
LR & Coefficient($\beta$) & 0.91 & 0.56 & 0.41 \\
  & Intercept($\alpha$) & 11.37 & 13.73 & 13.92 \\ \midrule
ARIMA & (p,d,q) & (1,1,0) & (1,1,0) & (1,1,0) \\
  & p-value & 0.90 & 0.96 & 0.94 \\
 & Min AIC & 4180 & 4714 & 2708 \\ \midrule
GA & Population \break Size & 10 & 10 & 10 \\
(Optimizer)  & Generation & 45 & 45 & 45 \\
  & Gene Length & 10 & 10 & 10\\
\bottomrule
\end{tabular}
\renewcommand{\arraystretch}{1}
\end{table}

\begin{table}[htb!]
\caption{Model Evaluation of CRD Mean Metrics}
\label{tab:model_evaluation_metrics}
\centering
\renewcommand{\arraystretch}{1}
\begin{tabular}{ l|rrr }
\toprule
\textbf{Models} & \textbf{RMSE} & \textbf{R\textsuperscript{2}} & \textbf{MAE} \\ 
\bottomrule
\toprule
LSTM & 15.19 & 0.52 & 103.67  \\
GRU & 4.10 & 0.47 & 71.03 \\ 
MLP & 17.22 & 0.67 & 818.67  \\ 
XGBR & 19.30 & 0.63 & 901.00 \\ 
RFR & 17.40 & 0.60 & 841.30 \\ 
SVR & 17.67 & 0.62 & 738.67 \\ 
LR & 19.60 & 0.62 & 868.01  \\ 
DTR & 37.96 & 0.24 & 1931.20 \\ 
AR & 51.39 & 0.23 & 2552.66  \\ 
ARIMA & 52.52 & 0.55 & 2686.00 \\ 
\bottomrule
\end{tabular}
\renewcommand{\arraystretch}{1}
\end{table}

\begin{figure}[htb!]
    \centering
    \begin{tikzpicture}[scale=0.6]

% \definecolor{color1}{rgb}{0.298039215686275,0.447058823529412,0.690196078431373}
\definecolor{dblue}{rgb}{0.29,0.44,0.69}

\begin{axis}[
legend cell align={left},
axis x line*=bottom,
axis y line*=right,
xtick pos=bottom,
ytick pos=left,
legend style={
at={(0.95,0.1)},anchor= east, nodes={scale=1},
draw=none
},
xbar, xmin=0,
width=12cm, height=10.0cm ,
xlabel={Rating}, ylabel={Model},
symbolic y coords={AR,ARIMA,DTR,RFR,SVR,XGBR,LR,MLP,GRU,LSTM},
ytick=data,
nodes near coords, nodes near coords align={horizontal},
cycle list name=exotic,
every axis plot/.append style={fill,draw=none,no markers}% <- added
]
\addplot [semithick, dblue, bar width=16pt] coordinates {(10,LSTM) (9,GRU)  (8,MLP) (7,LR) (6,XGBR) (5,SVR) (4,RFR) (3,DTR) (2,ARIMA) (1,AR) }; \label{dd};
\addlegendentry{Model Score}
\end{axis}
\end{tikzpicture}
    \caption{Performance Ranking of Models }
    \label{fig:surrogate_performance}
\end{figure}
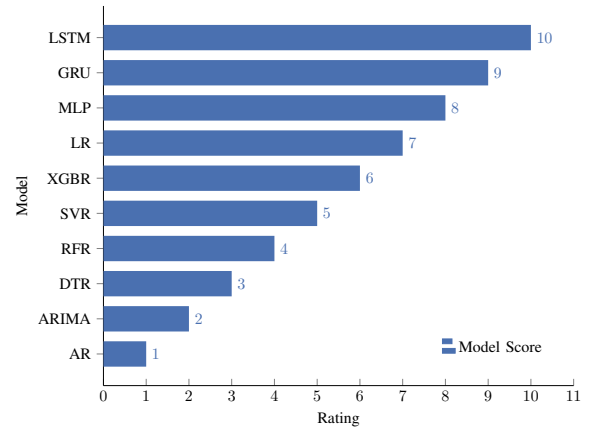

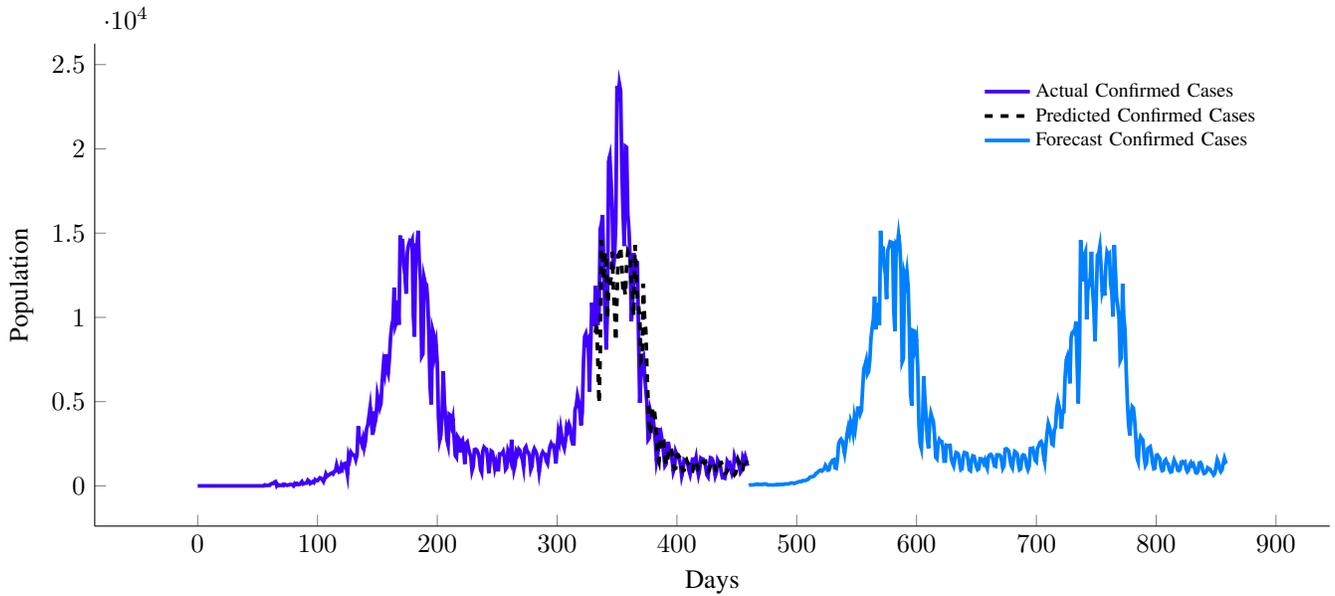
\begin{figure*}[htb!] %!t
% \centering
\begin{tikzpicture}[scale=1.0]

\definecolor{dblue}{rgb}{0.29,0.44,0.69}
\definecolor{dblue1}{rgb}{0.24998, 0, 1}
\definecolor{dblue2}{rgb}{0, 0.50002, 1}
\definecolor{dred}{rgb}{1.0,0.03,0.0}
\definecolor{dyellow}{rgb}{1.0, 0.65 0.0}
\definecolor{dgreen}{rgb}{0.0, 0.8,0.6}

\definecolor{dblack}{rgb}{0,0,0}

\begin{axis}
[
width=18cm,
height=8cm,
% scale only xaxis,
legend cell align={left},
axis x line*=bottom,
axis y line*=right,
xtick pos=bottom,
ytick pos=left,
legend style={
at={(0.95,0.85)},anchor=east,
draw=none, nodes={scale=0.75},
},
name = plot, xlabel={Days}, ylabel={Population},
]

\addplot [line width=1.4pt, dblue1]
table {./tikz/ctable_true_test.txt};
\addlegendentry{Actual Confirmed Cases}

\addplot [line width=1.4pt, dashed, dblack]
table {./tikz/ctable_prediction.txt};
\addlegendentry{Predicted Confirmed Cases}

\addplot [line width=1.4pt, dblue2]
table {./tikz/ctable_forecast.txt};
\addlegendentry{Forecast Confirmed Cases}

\end{axis}
\end{tikzpicture}
   
\caption{400-days COVID-19 Confirmed Cases Forecast in South Africa, (Starting 24-04-2021)}
\label{fig:forecast_confirmed_cases}
\end{figure*}

\begin{figure*}[htb!] %!t
\centering
\begin{tikzpicture}[scale=1.0]

\definecolor{dblue}{rgb}{0.29,0.44,0.69}
\definecolor{dred}{rgb}{1.0,0.03,0.0}
\definecolor{dyellow}{rgb}{1.0, 0.65 0.0}
\definecolor{dgreen1}{rgb}{0.0, 0.8,0.6} 
\definecolor{dgreen2}{rgb}{0, 1, 0.4}
\definecolor{dblack}{rgb}{0,0,0}

\begin{axis}
[
width=18cm,
height=8cm,
% scale only xaxis,
legend cell align={left},
axis x line*=bottom,
axis y line*=right,
xtick pos=bottom,
ytick pos=left,
legend style={
at={(0.95,0.85)},anchor=east,
draw=none, nodes={scale=0.75},
},
name = plot, xlabel={Days}, ylabel={Population},
]

\addplot [line width=1.4pt, dgreen1]
table {./tikz/dtable_true_test.txt};
\addlegendentry{Actual Recovered Cases}

\addplot [line width=1.4pt, dashed, dblack]
table {./tikz/dtable_prediction.txt};
\addlegendentry{Predicted Recovered Cases}

\addplot [line width=1.4pt, dgreen2]
table {./tikz/dtable_forecast.txt};
\addlegendentry{Forecast Recovered Cases}

\end{axis}
\end{tikzpicture}
   
\caption{400-days COVID-19 Recovered Cases Forecast in South Africa, (Starting 24-04-2021)}
\label{fig:forecast_recovered_cases}
\end{figure*}
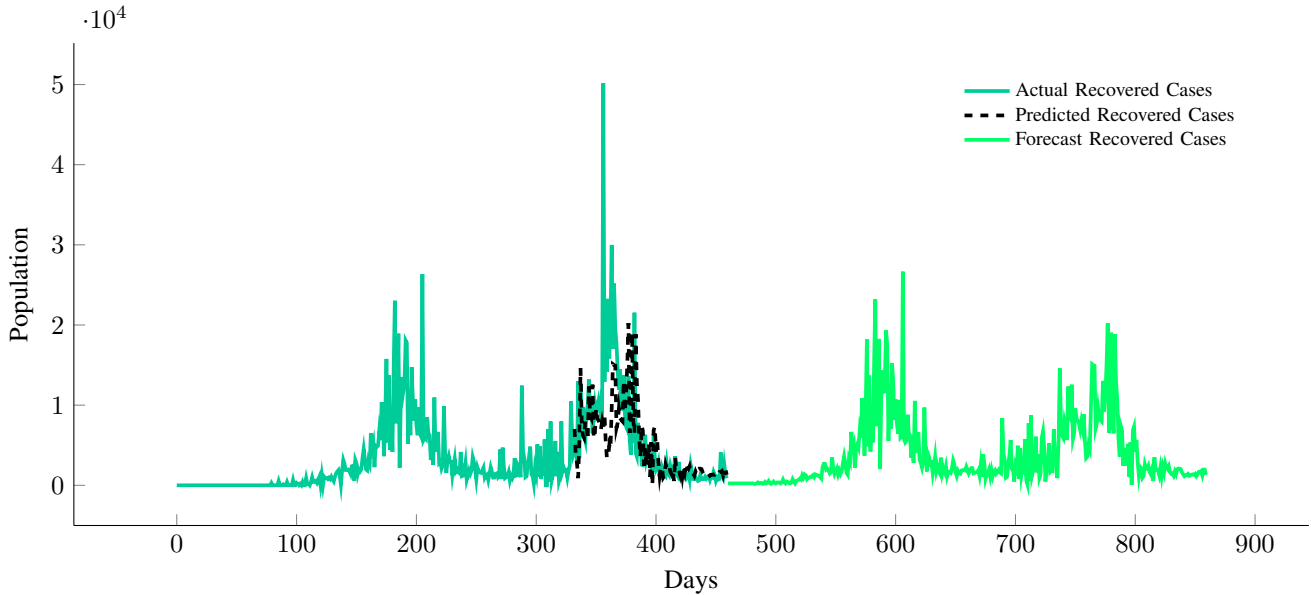

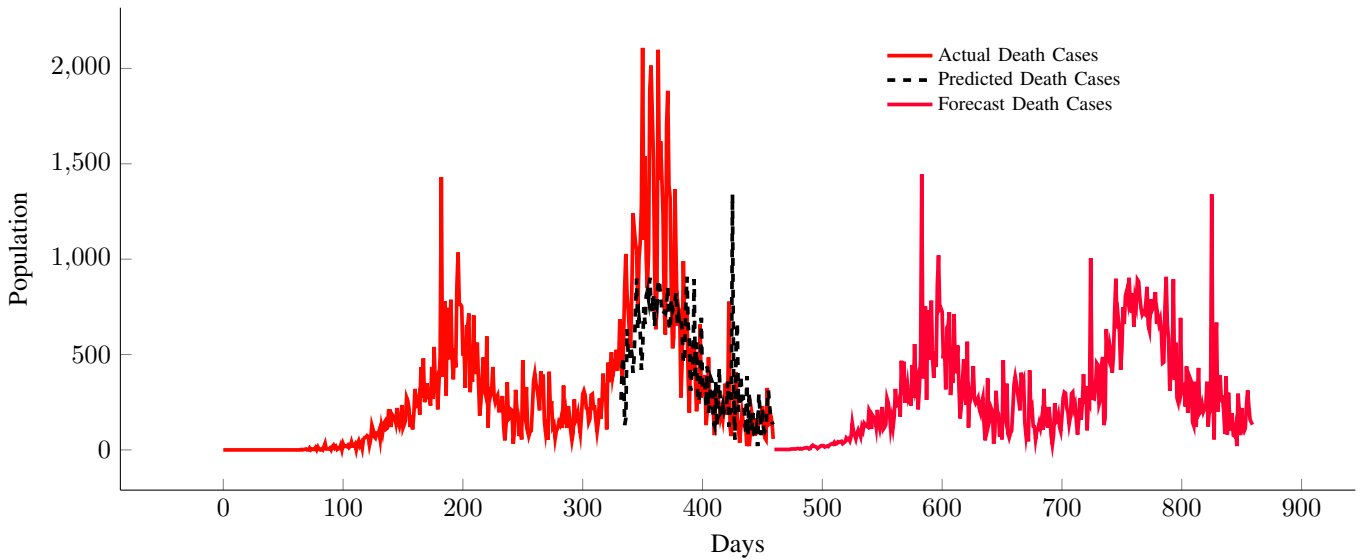
\begin{figure*}[htb!] %!t
\centering
\begin{tikzpicture}[scale=1.0]

\definecolor{dblue}{rgb}{0.29,0.44,0.69}
\definecolor{dred1}{rgb}{1.0,0.03,0.0}  
\definecolor{dred2}{rgb}{1, 0, 0.2}
\definecolor{dyellow}{rgb}{1.0, 0.65 0.0}
\definecolor{dgreen}{rgb}{0.0, 0.8,0.6}
\definecolor{dblack}{rgb}{0,0,0}

\begin{axis}
[
width=18cm,
height=8cm,
% scale only xaxis,
legend cell align={left},
axis x line*=bottom,
axis y line*=right,
xtick pos=bottom,
ytick pos=left,
legend style={
at={(0.82,0.85)},anchor=east,
draw=none, nodes={scale=0.75},
},
name = plot, xlabel={Days}, ylabel={Population},
]

\addplot [line width=1.4pt, dred1]
table {./tikz/etable_true_test.txt};
\addlegendentry{Actual Death Cases}

\addplot [line width=1.4pt, dashed, dblack]
table {./tikz/etable_prediction.txt};
\addlegendentry{Predicted Death Cases}

\addplot [line width=1.4pt, dred2]
table {./tikz/etable_forecast.txt};
\addlegendentry{Forecast Death Cases}

\end{axis}
\end{tikzpicture}
   
\caption{400-days COVID-19 Death Cases Forecast in South Africa, (Starting 24-04-2021)}
\label{fig:forecast_death_cases}
\end{figure*}

As shown in Table~\ref{tab:hyperparameter_selection}, our study demonstrates the process of obtaining an optimal hyperparameter value(s) for each of the models that we implemented. 
The cost function uses the GA to identify surrogate hyperparameters by minimizing the RMSE during the search process and maximizing accuracy. The mutation rate and crossover function were defined in the architecture of the GA.

In Table~\ref{tab:hyperparameter_selection}, the heading labeled as \textbf{C}, \textbf{R}, and \textbf{D} represents the confirmed, recovered and death cases respectively.

Table~\ref{tab:model_evaluation_metrics} shows the evaluation metrics of the models considered in our study. Likewise, Figure~\ref{fig:surrogate_performance} illustrates the performance rank of the models explored in this study, with a rating of 1 to 10, with 10 being the best performing rank.

The 400-days projection by the benchmark surrogate (LSTM) showing the transmission trend for confirmed, recovered, and death cases of COVID-19 in South Africa from April $24^{th}$ 2021 till May $22^{nd}$ 2022 are plotted in Figure~\ref{fig:forecast_confirmed_cases}, Figure~\ref{fig:forecast_recovered_cases} and Figure~\ref{fig:forecast_death_cases} respectively. 

\section{Results and Discussion}\label{result_and_discussion}
Our research is a univariate time series analysis of the COVID-19 datasets of confirmed, recovered, and death cases in South Africa. We ensured that the time-series data are stationary for surrogate simulation as described in Section~\ref{methods_models}. We used 70\% of the CRD dataset for training the surrogate, while 30\% was used for testing. The results from our research are discussed further in the sub-sections below.

Relating to the third wave of the COVID-19 in South Africa, Figure~\ref{fig:forecast_confirmed_cases} shows the transmission dynamics of the confirmed cases, Figure~\ref{fig:forecast_recovered_cases} shows the trend of disease recovery by infected individuals, and Figure~\ref{fig:forecast_death_cases} represents the predicted death pattern.

We discovered that LSTM can remember longer data sequences than GRU, although, GRU has less hyperparameter complexity and is easier to analyze. The relationship between the SEIR model, DES, and surrogate is that, the surrogate offers future predictions based on real-world data, and we use the SEIR results to group a random population sample into epidemiological compartments. Likewise, the DES retrieves the number infectious individuals in the SEIR model as the starting population for its event simulation.

We predict the future trend of COVID-19 CRD cases up to a specified duration $n=400$. This study does not consider the impact of hospital resource distribution, economic effect, or the administration of personal protective equipment (PPE) required by healthcare professionals.

Our work is significant because it addresses the development of a system with an optimal function capable of making an approximate daily forecast of the pandemic instance in South Africa. In addition, the prediction covers the projected third wave of COVID-19 in South Africa.

% ==================
% # IV. CONCLUSION #
% ==================

\section{Conclusion} \label{conclusion}
By comparing various machine learning models and statistical models for predicting future transmission trend of COVID-19 in South Africa, we observed that the deep learning models outperform the conventional and statistical models as illustrated in Figure~\ref{fig:surrogate_performance}. We also demonstrated that our optimized surrogate can forecast the transmission dynamics of COVID-19 or other pandemic disease in the future with the same data presentation and features.

The data pattern from the COVID-19 CRD datasets indicates that significant disease spread can be controlled if South African government authorities adopt timely and functional preventive action at the emergence of the disease.

Following model simulation and data fitting, our surrogate forecasted COVID-19 in South Africa for 400-days (beginning April 24th, 2021). The projection is based on the disease's historical transmission pattern during the last 458-days (January $22^{nd}$, 2020 - April $23^{rd}$, 2021). The long-term forecast captures the anticipated third-wave occurrence of the COVID-19 pandemic in South Africa.

This study has demonstrated the impact of surrogate models for forecasting pandemic disease for range $[0,n]$. The upper range $n$ is advised to be less or equal to 500, i.e. $n\leq 500$ to avoid skew forecast or early convergence of results. This demonstrates that ADRIANA's result can guide decisions for adequate resources planning during a pandemic. 

For future research, the ADRIANA system can be improved to respond dynamically to different pandemic cases while taking into consideration external factors (such as geolocation, the effect of weather, government policy type, and vaccination distribution to the population), simulate hospital resources for treatment, simulate the allocation of personal protective equipment (PPE) and healthcare personnel to hospitals to allow for more model robustness and flexibility. 

For our results to be reproducible, the source file for engineering the ADRIANA system can be accessed through \href{https://github.com/timzzy/adriana}{ this link}.

% ==================
% # ACKNOLEDGMENTS #
% ==================

\section*{Acknowledgment}
The authors would like to appreciate the Computer Science department at the University of Witwatersrand, and Institute of Intelligent Systems at the University of Johannesburg, South Africa for the interest in providing a computational solution to the COVID-19 outbreak in South Africa. This research is supported by the Nedbank Research Chair. 

% ==============
% # REFERENCES #
% ==============

\newpage

\bibliographystyle{IEEEtran}
\bibliography{tim}

\begin{thebibliography}{10}
\providecommand{\url}[1]{#1}
\csname url@samestyle\endcsname
\providecommand{\newblock}{\relax}
\providecommand{\bibinfo}[2]{#2}
\providecommand{\BIBentrySTDinterwordspacing}{\spaceskip=0pt\relax}
\providecommand{\BIBentryALTinterwordstretchfactor}{4}
\providecommand{\BIBentryALTinterwordspacing}{\spaceskip=\fontdimen2\font plus
\BIBentryALTinterwordstretchfactor\fontdimen3\font minus
  \fontdimen4\font\relax}
\providecommand{\BIBforeignlanguage}[2]{{%
\expandafter\ifx\csname l@#1\endcsname\relax
\typeout{** WARNING: IEEEtran.bst: No hyphenation pattern has been}%
\typeout{** loaded for the language `#1'. Using the pattern for}%
\typeout{** the default language instead.}%
\else
\language=\csname l@#1\endcsname
\fi
#2}}
\providecommand{\BIBdecl}{\relax}
\BIBdecl

\bibitem{vanzyl2021did}
Terence,~V.~Z. and Turgay,~C., ``Did we produce more waste during the covid-19
  lockdowns a remote sensing approach to landfill change analysis,'' \emph{IEEE
  Journal of Selected Topics in Applied Earth Observations and Remote Sensing},
  2021.

\bibitem{cite1}
Wang,~H. and Dong,~Z., ``Phase-adjusted estimation of the number of coronavirus
  disease 2019 cases in wuhan, china.'' \emph{Cell Discov}, vol.~47, no.~8, pp.
  1680--1681, October 2000.

\bibitem{cite2}
Chaolin,~C., B.Huang, and Yeming,~M., ``Clinical features of patients infected
  with 2019 novel coronavirus in wuhan, china,'' \emph{The Lancet}, vol. 395,
  no. 10223, pp. 497--506, February 2020.

\bibitem{cite38}
Aleman,~D.~M., Wibisono,~T.~G., and Schwartz,~B., ``A non-homogeneous agent
  based simulation approach to modeling the spread of disease in a pandemic
  outbreak,'' \emph{Interfaces}, vol.~41, no.~3, pp. 215--301, May 2011.

\bibitem{cite39}
Rylan,~P. and van Zyl Terence~L, ``Surrogate assisted methods for the
  parameterisation of agent-based models,'' in \emph{7th International
  Conference on Soft Computing Machine Intelligence (ISCMI)}, November 2020,
  pp. 78--82.

\bibitem{stander2020extended}
L,~S., M,~W., van Zyl, and TL, ``Extended surrogate assisted continuous process
  optimisation,'' in \emph{2020 7th International Conference on Soft Computing
  \& Machine Intelligence (ISCMI)}.\hskip 1em plus 0.5em minus 0.4em\relax
  IEEE, 2020, pp. 275--279.

\bibitem{cite3}
Thengade,~O., Anita,~C., and Rucha,~D., ``Genetic algorithm – survey paper,''
  \emph{IJCA Proc National Conference on Recent Trends in Computing}, vol.~5,
  no. 123, pp. 499--508, January 2012.

\bibitem{cite5}
Martínez,~S.~Z. and Coello,~C.~A., ``Moea/d assisted by rbf networks for
  expensive multi-objective optimization problems,'' \emph{Proc. ACM Genet.
  Evol. Comput. Conf., Amsterdam, The Netherlands}, vol.~8, p. 1405–1412.,
  March 2013.

\bibitem{cite7}
Jin,~Y., Olhofer,~M., and Sendhoff,~B., ``A framework for evolutionary
  optimization with approximate fitness function,'' \emph{IEEE Trans. Evol.
  Comput.}, vol.~6, no.~5, p. 481–494, October 2002.

\bibitem{cite8}
Chugh,~T., Jin,~Y., Miettinen,~K., Hakanen,~J., and Sindhya,~K., ``A
  surrogate-assisted reference vector guided evolutionary algorithm for
  computationally expensive many-objective optimization,'' \emph{IEEE Trans.
  Evol. Comput.}, vol.~22, no.~1, pp. 129–142,, 02 2018.

\bibitem{cite9}
Buche,~D., Schraudolph,~N.~N., and Koumoutsakos,~P., ``Accelerating
  evolutionary algorithms with gaussian process fitness function models,''
  \emph{IEEE Trans. Syst.}, vol.~35, no.~2, p. 183–194, May 2005.

\bibitem{cite10}
Zhou,~Z., Ong,~Y.~S., Nguyen,~M.~H., and Lim,~D., ``A study on polynomial
  regression and gaussian process global surrogate model in hierarchical
  surrogate-assisted evolutionary algorithm,'' \emph{Proc. IEEE Congr. Evol.
  Comput. (CEC), Edingburg, UK}, vol.~3, no.~3, p. 2832–2839, September 2005.

\bibitem{cite20}
Douglas,~B. and Bernard,~L., ``A framework for rationing ventilators and
  critical care beds during the covid-19 pandemic,'' pp. 1773--1774, March
  2020.

\bibitem{cite22}
Ashleigh,~R., Amy,~L., Michael,~W., Anne-Luise,~W., Brenda,~L., Yan,~P.,
  Wu,~J., Seyed,~M., David,~B., Babak,~P., and David,~N., ``Estimated
  epidemiologic parameters and morbidity associated with pandemic h1n1
  influenza,'' pp. 1773--1774, March 2020.

\bibitem{cite14}
Liestol,~K. and Andersen,~P., ``Updating of covariates and choice of time
  origin in survival analysis: problems with vaguely defined disease states,''
  \emph{Stat Med}, vol.~21, no.~23, pp. 3701--3714, June 2002.

\bibitem{cite15}
Kratschmer,~V., ``Strong consistency of least-squares estimation in linear
  regression models with vague concepts,'' \emph{J Multivar Anal}, vol.~97, pp.
  633--654, March 2006.

\bibitem{mathonsi2020prediction}
Thabang,~M. and van Zyl Terence~L, ``Prediction interval construction for
  multivariate point forecasts using deep learning,'' in \emph{2020 7th
  International Conference on Soft Computing \& Machine Intelligence
  (ISCMI)}.\hskip 1em plus 0.5em minus 0.4em\relax IEEE, 2020, pp. 88--95.

\bibitem{cite11}
\BIBentryALTinterwordspacing
Benvenuto,~D., Giovanetti,~M., Vassallo,~L., Angeletti,~S., and Ciccozzi,~M.,
  ``Application of the arima model on the covid-2019 epidemic dataset,''
  \emph{Proc. IEEE Congr. Evol. Comput. (CEC), Edingburg, UK}, p. 105340, 07
  2020. [Online]. Available: \url{https://doi.org/10.1016/j.dib.2020.105340}
\BIBentrySTDinterwordspacing

\bibitem{cite12}
\BIBentryALTinterwordspacing
Dehesh,~T., Mardani-Fard,~H., and Dehesh,~P., ``Forecasting of covid-19
  confirmed cases in different countries with arima models,'' \emph{MedRxiv},
  p. 105340, 06 2020. [Online]. Available:
  \url{https://doi.org/10.1101/2020.03.13.20035345}
\BIBentrySTDinterwordspacing

\bibitem{cite13}
Wang,~Y., Shen,~Z., and Jiang,~Y., ``Comparison of arima and gm(1,1) models for
  prediction of hepatitis b in china,'' \emph{PloS One}, vol.~13, September
  2018.

\bibitem{cite16}
Shawni,~D., Bandyopadhyay, and Kumar,~S., ``Machine learning approach for
  confirmation of covid-19 cases: positive, negative, death and release,''
  \emph{medRxiv Journal}, March April 2020.

\bibitem{cite17}
Huang,~C., Chen,~Y., Ma,~Y., and Kuo,~P., ``Multiple-input deep convolutional
  neural network model for covid-19 forecasting in china,'' \emph{medRxiv
  journal}, vol.~30, no.~5, March 2020.

\bibitem{cite18}
Gary,~E., Weissman,~M., Crane-Droesch,~A., and Chivers,~C., ``Locally informed
  simulation to predict hospital capacity needs during the covid-19 pandemic,''
  \emph{Annals of internal medicine}, vol. 173, no.~1, pp. 21--28, April 2020.

\bibitem{cite19}
Schaar,~V.~D., ``Adjutorium: Using covid-19 patient data to train machine
  learning models for healthcare,'' \emph{Journal of Health Informatics:
  University of Cambridge}, May 2020.

\bibitem{cite21}
Aggarwal,~B. and Anupama,~J., ``Epidemic simulation model,'' December 2019.

\bibitem{cite23}
Kumar,~P., Chimmula,~R., and Zhang,~L., ``Time series forecasting of covid-19
  transmission in canada using lstm networks,'' pp. 109\,864--109\,864, June
  2020.

\bibitem{cite24}
Jophn,~H., ``Covid-19 dataset,''
  \url{https://github.com/CSSEGISandData/COVID-19}, February 2020.

\bibitem{cite25}
Bisgaard,~S. and Kulahci,~M., \emph{Time Series Analysis and Forecasting by
  Example}.\hskip 1em plus 0.5em minus 0.4em\relax John Willey and Sons. Inc.,
  2011.

\bibitem{cite28}
\BIBentryALTinterwordspacing
Holmes,~E., Scheuerell,~M.~D., and Ward,~E.~J., ``Applied time series analysis
  for fisheries and environmental sciences.'' [Online]. Available:
  \url{https://nwfsc-timeseries.github.io/atsa-labs/sec-boxjenkins-aug-dickey-fuller.html}
\BIBentrySTDinterwordspacing

\bibitem{cite29}
Kermack,~W.~O. and McKendrick,~A.~G., ``A contribution to the mathematical
  theory of epidemics,'' \emph{Proc. R. Soc. London}, vol. 115, no. 772, pp.
  700--721, August 1927.

\bibitem{cite30}
Hutzenthaler,~M., Jentzen,~A., and Kloeden,~P.~E., ``Strong and weak divergence
  in finite time of euler's method for stochastic differential equations with
  non-globally lipschitz continuous coefficients,'' \emph{Proc. R. Soc.}, vol.
  467, no. 2130, pp. 1563--15\,767, June 1927.

\bibitem{cite40}
Fujimoto,~R.~M., ``Parallel discrete event simulation,'' \emph{Communications
  of the ACM}, vol.~33, no.~10, pp. 30--53, October 1990.

\bibitem{cite31}
Li.,~H., Li,~W., Cook,~C., Zhu,~C., and Gao,~Y., ``Independently recurrent
  neural network (indrnn): Building a longer and deeper rnn,'' \emph{IEEE
  Conference on Computer Vision and Pattern Recognition (CVPR)}, pp.
  5457--5466, June 2018.

\bibitem{cite32}
Donkers,~T., Loepp,~B., and Ziegler,~J., ``Sequential user-based recurrent
  neural network recommendations,'' \emph{Proceedings of the Eleventh ACM
  Conference on Recommender Systems}, p. 152–160, August 2017.

\bibitem{cite34}
Engelbrecht,~A.~P., \emph{Computational Intelligence: An Introduction, 2nd
  Edition}.\hskip 1em plus 0.5em minus 0.4em\relax Addison-Wesley, 2007.

\bibitem{cite35}
Krause,~P., Boyle,~D.~P., and Base,~F., ``Comparison of different efficiency
  criteria for hydrological model assessment,'' \emph{Advances in Geosciences},
  vol.~5, pp. 89--97, October 2005.

\bibitem{cite36}
Barnston,~A., ``Correspondence among the correlation [root mean square error]
  and heidke verification measures; refinement of the heidke score,''
  \emph{Climate Analysis Center}, vol.~8, July 1992.

\bibitem{cite4}
Metawa,~N., Hassan,~K., and Elhoseny,~M., ``Genetic algorithm based model for
  optimizing bank lending decisions,'' \emph{The Expert Systems with
  Applications}, vol.~80, no. 10223, pp. 75--82, March 2017.

\end{thebibliography}

% \bibliography{IEEEabrv,tim}
\end{document}